\documentclass[oneside,english]{amsart}
\usepackage[T1]{fontenc}
\usepackage[left=1in, right=1in, top=1in, bottom=1in]{geometry}
\usepackage[latin9]{inputenc}
\usepackage{amsthm}
\usepackage{amssymb}
\usepackage{amsmath}
\usepackage{amsaddr}
\usepackage[normalem]{ulem}
\usepackage{natbib}
\usepackage{setspace}

\makeatletter
\numberwithin{equation}{section}
\numberwithin{figure}{section}
\theoremstyle{plain}
\newtheorem{lem}{Lemma}
  \theoremstyle{plain}

\makeatother

\usepackage{babel}

\begin{document}

\title{A note on breaking ties among sample medians}

\author{Peter M. Aronow and Donald K.K. Lee}

\address{Yale University and Emory University}


\maketitle
\doublespacing

\begin{abstract}
Given samples $x_1,\cdots,x_n$, it is well known that any sample median value (not necessarily unique) minimizes the absolute loss $\sum_{i=1}^n |q-x_i|$. Interestingly, we show that the minimizer of the loss $\sum_{i=1}^n|q-x_i|^{1+\epsilon}$ exhibits a singular perturbation behaviour that provides a unique definition for the sample median as $\epsilon \rightarrow 0$. This definition is the unique point among all candidate median values that balances the {\it logarithmic} moment of the empirical distribution. The result generalizes directly to breaking ties among sample quantiles when the quantile regression loss is modified in the same way.
\end{abstract}

\section{Introduction}

Given samples $x_1,\cdots,x_n$, it is well known that
the sample mean $n^{-1}\sum_i x_i$ is the unique minimizer of the empirical squared loss $\mathbb{E}_{n}(\theta-X)^{2}=n^{-1}\sum_{i:x_i\le\theta}(\theta-x_i)^{2}+n^{-1}\sum_{i:x_i>\theta}(x_i-\theta)^{2}$.
This follows from the first order condition
\[
n^{-1}\sum_{i:x_i\le\theta}(\theta-x_i) = n^{-1}\sum_{i:x_i>\theta}(x_i-\theta),
\]
which can be seen as finding the point $\theta$ that balances the
first moment of the distribution. 

It is also well known that the sample median need not be unique, but can take on an interval of values if $n$ is even. If it is the
absolute loss $\mathbb{E}_{n}|\theta-X|=n^{-1}\sum_{i:x_i\le\theta}(\theta-x_i)+n^{-1}\sum_{i:x_i>\theta}(x_i-\theta)$
that one is interested in minimizing, then any median value satisfying
$F_n(\theta)=1/2$ (where $F_n(x) = n^{-1}\sum_i I(x_i \le x)$ is the empirical distribution) is a solution to the first order condition\footnote{Since $\sum_{i:x_i\leq\theta}(\theta-x_i)$ has a subderivative whenever $\theta=x_i$, $(\theta-x_i)^0$ in the first order condition \eqref{eq:0-moment} is allowed to take on any value in the interval $[0,1]$ when $\theta=x_i$.}
\begin{equation}
\label{eq:0-moment}
\underbrace{n^{-1}\sum_{i:x_i\le\theta} (\theta-x_i)^0}_{F_n(\theta)} = \underbrace{n^{-1}\sum_{i:x_i>\theta} (x_i-\theta)^0}_{1-F_n(\theta)},
\end{equation}
which seeks any point that balances the zero-th moment of the empirical distribution. Informally, the non-uniqueness of the median can be attributed to
the fact that merely balancing the zero-th moment does not provide enough `discriminative' power, while balancing the first moment does. 

In order to report a unique sample median, some method of breaking ties among candidate median values is necessary. Textbook treatments and software packages typically define the sample median as the midpoint of the interval \citep{hf}.  Equivalent problems emerge in the calculations of sample quantiles in general. A variety of alternative estimators based on interpolation, linear combinations of order statistics, or smoothing-type approaches \citep{hd,parrish,sv,sm,yang} have been proposed, typically under the assumption of IID samples from a population with a uniquely defined quantile (e.g., when the population distribution is continuous). 

In this note, we show that balancing an ever so slightly higher order moment than the zero-th one leads to a way to tiebreak among the sample medians. Recalling that $\log x$ is asymptotically dominated by $x^p$ for any $p>0$, consider choosing $\theta$ to balance the \textit{logarithmic} moment:
\begin{equation}
\sum_{i:x_i<\theta}\log(\theta-x_i) = \sum_{i:x_i>\theta}\log(x_i-\theta).\label{eq:log-moment}
\end{equation}
We show that this is equivalent to the minimization of $\mathbb{E}_{n}|\theta-X|^{1+\epsilon}$ in the limit $\epsilon\downarrow0$: The unique minimizer of $\mathbb{E}_{n}|\theta-X|^{1+\epsilon}$ converges to a candidate value for the median as $\epsilon \downarrow 0$. If there are multiple candidate values, then the one that balances (\ref{eq:log-moment}) is the unique limit. This singular perturbation behaviour of the first order condition converging to (\ref{eq:log-moment}) rather than to (\ref{eq:0-moment}) gives rise to an interesting way for defining the median uniquely. The idea generalizes directly to defining unique sample quantiles $q_{\alpha}$ when the quantile regression loss is modified in the same way.

%
%
%

\section{Result}

Given $\alpha \in (0,1)$, we define a modified version of the weighted absolute loss
for quantile regression as
\begin{equation}\label{eq:loss}
L_{\alpha,\epsilon}(x,q)=\begin{cases}
(1-\alpha)(q-x)^{1+\epsilon} & x\leq q\\
\alpha(x-q)^{1+\epsilon} & x> q
\end{cases}.
\end{equation}
If $\epsilon=0$ then we have the usual loss used in quantile regression,
whose expectation with respect to an empirical distribution $F_{n}(x)$
is minimized by any $\alpha$-quantile $q_{\alpha}$ satisfying $F_{n}(q_{\alpha})=\alpha$. The median naturally corresponds to the case where $\alpha = 1/2$.

The expectation of $L_{\alpha,\epsilon}(x,q)$ is
\begin{equation}
\mathbb{E}_{n}L_{\alpha,\epsilon}(x,q)=\frac{1-\alpha}{n}\sum_{i:x_i\le q}(q-x_i)^{1+\epsilon}+\frac{\alpha}{n}\sum_{i:x_i> q}(x_i-q)^{1+\epsilon},\label{eq:E-loss}
\end{equation}
and its derivative at $q$ is
\begin{equation}\label{eq:gradient}
\frac{1-\alpha}{n}\sum_{i:x_i\le q}(q-x_i)^\epsilon-\frac{\alpha}{n}\sum_{i:x_i> q}(x_i-q)^\epsilon
\end{equation}
up to a factor $1+\epsilon$.

When $\epsilon>0$, $\mathbb{E}_{n}L_{\alpha,\epsilon}(x,q)$ has a unique minimizer because it is strongly convex in $q$. The minimizer balances the weighted $\epsilon$-th order moment in (\ref{eq:gradient}). Whereas for $\epsilon=0$ the zero-th order moment is balanced by possibly many values. Lemma \ref{lem:main} below shows that the minimization of $\mathbb{E}_{n}L_{\alpha,\epsilon}(x,q)$ as $\epsilon\downarrow0$ is qualitatively very different from the minimization of $\mathbb{E}_{n}L_{\alpha,0}(x,q)$.

\vspace{0.1in}

\begin{lem}\label{lem:main} Let $q_{\alpha,\epsilon}$ be the minimizer of $\mathbb{E}_{n}L_{\alpha,\epsilon}(x,q)$.

\noindent (i) Suppose there exists a unique $\alpha$-quantile $q_{\alpha}$, i.e. $F_{n}(q_{\alpha}-)<\alpha$ and $F_{n}(q_{\alpha})>\alpha$.
Then it is the limit of $q_{\alpha,\epsilon}$ as $\epsilon\downarrow 0$.

\noindent (ii) If no unique $\alpha$-quantile
exists, then $F_{n}(q)=\alpha$ in some interval $[q_{\alpha}^{L},q_{\alpha}^{H})$.
The unique solution $q^{\log}_{\alpha} \in (q_{\alpha}^{L},q_{\alpha}^{H})$ that balances
the weighted log-moment
\begin{equation}
(1-\alpha)\sum_{i:x_i< q}\log(q-x_i)-\alpha\sum_{i:x_i> q}\log(x_i-q)=0\label{eq:wlog-moment}
\end{equation}
is the limit of $q_{\alpha,\epsilon}$ as $\epsilon\downarrow 0$.
\end{lem}

The intuition for the result is simple but elegant: Perturbing $\epsilon$ about 0 yields approximations for the terms
\[
n^{-1}\sum_{i:x_i\le q}(q-x_i)^{\epsilon} \approx F_n(q-)+\frac{\epsilon}{n}\sum_{i:x_i< q}\log(q-x_i),
\]
\[
n^{-1}\sum_{i:x_i > q}(x_i-q)^{\epsilon} \approx 1-F_n(q)+\frac{\epsilon}{n}\sum_{i:x_i > q}\log(x_i-q).
\]
Ignoring differences between $F_n(q-)$ and $F_n(q)$ for a moment, the first order condition obtained from setting the derivative (\ref{eq:gradient}) to zero is
\[
F_{n}(q)-\alpha
+\frac{\epsilon}{n}\left\{ (1-\alpha)\sum_{i:x_i\le q}\log(q-x_i) - \alpha\sum_{i:x_i>q}\log(x_i-q)\right\} \approx 0.
\]
The dominant term above is $F_n(q)-\alpha$, so the limiting minimizer has to be an $\alpha$-quantile. Among the candidate $\alpha$-quantiles $[q_\alpha^L,q_\alpha^H)$ in case (ii), the term in the curly brackets now become dominant, giving rise to the logarithmic moment condition (\ref{eq:wlog-moment}).

\begin{proof}
For case (i) where there is a unique $\alpha$-quantile $q_{\alpha}$ (at the location of one of the samples $x_i$), set $q=q_{\alpha}-\delta$ for a small $\delta>0$ and use Taylor's theorem to obtain
\[
n^{-1}\sum_{i:x_i\le q}(q-x_i)^{\epsilon} =
F_{n}(q_{\alpha}-)+\mathcal{O}(\epsilon), 
\]
\[
n^{-1}\sum_{i:x_i > q}(x_i-q)^{\epsilon} = 1-F_{n}(q_{\alpha}-)+\mathcal{O}(\epsilon).
\]
The derivative (\ref{eq:gradient}) at $q=q_{\alpha}-\delta$ is then $F_{n}(q_{\alpha}-)-\alpha+\mathcal{O}(\epsilon)<0$
for $\epsilon$ small enough. Likewise, the derivative at $q=q_{\alpha}+\delta$
is $F_{n}(q_{\alpha})-\alpha+\mathcal{O}(\epsilon)>0$ for $\epsilon$
small enough. Given that $\mathbb{E}_{n}L_{\alpha,\epsilon}(x,q)$ is strongly convex, its minimizer $q_{\alpha,\epsilon}$ must then be within $(q_\alpha-\delta,q_\alpha+\delta)$ for $\epsilon$ sufficiently small.

For case (ii), note that $F_{n}(x)$ has atoms at $x=q_{\alpha}^{L},q_{\alpha}^{H}$
but none in $(q_{\alpha}^{L},q_{\alpha}^{H})$. Hence within this
interval, the first sum in (\ref{eq:wlog-moment}) is increasing
in $q$ while the second one is decreasing. Moreover the left hand
side of (\ref{eq:wlog-moment}) approaches $-\infty$ as $q\downarrow q_{\alpha}^{L}$,
and approaches $+\infty$ as $q\uparrow q_{\alpha}^{H}$. Hence (\ref{eq:wlog-moment})
has a unique solution $q_{\alpha}^{\log}$ in $(q_{\alpha}^{L},q_{\alpha}^{H})$. Within
this interval, applying Taylor's theorem shows that
\[
n^{-1}\sum_{i:x_i\le q}(q-x_i)^{\epsilon} = \alpha +\frac{\epsilon}{n}\sum_{i:x_i< q}\log(q-x_i) + \mathcal{O}(\epsilon^{2}),
\]
\[
n^{-1}\sum_{i:x_i > q}(x_i-q)^{\epsilon}=1-\alpha+\frac{\epsilon}{n}\sum_{i:x_i > q}\log(x_i-q)+\mathcal{O}(\epsilon^{2}),
\]
so the derivative \eqref{eq:gradient} is
\[
(1-\alpha)\sum_{i:x_i< q}\log(q-x_i)-\alpha\sum_{i:x_i > q}\log(x_i-q) + \mathcal{O}(\epsilon)
\]
up to a factor $\epsilon/n$. For a small $\delta>0$, the derivative of $\mathbb{E}_{n}L_{\alpha,\epsilon}(x,q)$ at $q = q_{\alpha}^{\log} - \delta$ is then negative for a sufficiently small $\epsilon$, and likewise the derivative at $q = q_{\alpha}^{\log} + \delta$ is positive for $\epsilon$ small enough. The result then follows from the same line of argument for case (i).

\end{proof}

\section{Discussion}

This note serves to show that the introduction of a homotopy between the squared loss (which has a unique minimizer) and the absolute loss (which may have multiple minimizers) can be a means for resolving the non-uniqueness of the sample median. Our result may have implications for a broader family of problems, including non-uniqueness issues that arise in least absolute deviations regression and in quantile regression. While conceptual in value, our result provides insight into a canonical statistical problem and may spur practical innovations in future work.

%

\end{document}